\documentclass[aps,twocolumn,superscriptaddress]{revtex4}

\usepackage{graphicx}

\begin{document}
\bibliographystyle{revtex}
\draft


\title{Field-induced magnetic order in La$_{2-x}$Sr$_x$CuO$_4$ ($x =$ 0.10, 0.115, 0.13) studied by in-plane thermal conductivity measurements}



\author{K. Kudo}
\thanks{Present address: Institute for Materials Research (IMR), Tohoku University, Katahira 2-1-1, Aoba-ku, Sendai 980-8577, Japan}
\email{kudo@imr.tohoku.ac.jp}
\author{M. Yamazaki}
\author{T. Kawamata}
\author{T. Adachi}
\author{T. Noji}
\author{Y. Koike}
\affiliation{Department of Applied Physics, Graduate School of Engineering, Tohoku University, Aoba-yama 05, Aoba-ku, Sendai 980-8579, Japan}
\author{T. Nishizaki}
\author{N. Kobayashi}
\affiliation{Institute for Materials Research (IMR), Tohoku University, Katahira 2-1-1, Aoba-ku, Sendai 980-8577, Japan}


\date{November 17, 2003}

\begin{abstract}
 We have measured the thermal conductivity in the $ab$-plane of La$_{2-x}$Sr$_x$CuO$_4$ ($x =$ 0.10, 0.115, 0.13) in magnetic fields up to 14 T parallel to  the $c$-axis and also parallel to the $ab$-plane.
By the application of magnetic fields parallel to the $c$-axis, the thermal conductivity has been found to be suppressed at low temperatures below the temperature $T_\kappa$ which is located above the superconducting transition temperature and is almost independent of the magnitude of the magnetic field. 
The suppression is marked in $x =$ 0.10 and 0.13, while it is small in $x =$ 0.115. 
Furthermore, no suppression is observed in the 1 \% Zn-substituted La$_{2-x}$Sr$_x$Cu$_{0.99}$Zn$_{0.01}$O$_4$ with $x =$ 0.115. 
Taking into account the experimental results that the temperature dependence of the relative reduction of the thermal conductivity  is quite similar to the temperature dependence of the intensity of the incommensurate magnetic Bragg peak corresponding to the static stripe order and that the Zn substitution tends to stabilize the static order, it is concluded that the suppression of the thermal conductivity in magnetic fields is attributed to the development of the static stripe order. 
The present results suggest that the field-induced magnetic order in La$_{2-x}$Sr$_x$CuO$_4$ originates from the pinning of the dynamical stripes of spins and holes by vortex cores.
\end{abstract}

\maketitle

\section{Introduction}
The so-called 1/8 anomaly, namely, the anomalous suppression of superconductivity at $p$ (the hole concentration per Cu in the CuO$_2$ plane) $\sim$ 1/8 in the La-based high-$T_{\rm c}$ superconductors has been understood in terms of the stripe model proposed by Tranquada {\it et al.}\cite{rf:Tranquada}. 
From their neutron scattering experiment, a static stripe order of spins and holes was suggested to be formed at low temperatures in La$_{1.48}$Nd$_{0.4}$Sr$_{0.12}$CuO$_{4}$.  
Meanwhile, a spatially modulated dynamical spin correlation has been found to exist in a wide range of $p$ from the underdoped to overdoped region in La$_{2-x}$Sr$_x$CuO$_4$\cite{rf:Cheong,rf:Mason,rf:Thurston,rf:Yamada}.
Since the dynamical spin correlation may be regarded as a spin part of the dynamical stripe correlations of spins and holes, it has been understood that the dynamical stripes tend to be statically stabilized around $p =$ 1/8 and are easily pinned, leading to the static stripe order and the 1/8 anomaly\cite{rf:Tranquada,rf:Kumagai,rf:Luke,rf:Watanabe,rf:Watanabe2,rf:Koike0,rf:Koike,rf:Adachi}. 
For the pinning, the buckling of the CuO$_2$ plane in the tetragonal low-temperature (TLT) structure (space group: P4$_2$/ncm) is regarded as being effective in La$_{1.6-x}$Nd$_{0.4}$Sr$_x$CuO$_4$ or La$_{2-x}$Ba$_x$CuO$_4$. 
Moreover, a small amount of nonmagnetic impurities such as Zn has also been found to be effective for the pinning from the transport measurements\cite{rf:Koike0,rf:Koike,rf:Adachi} and also from the muon-spin-relaxation ($\mu$SR) measurements\cite{rf:muSR0,rf:muSR1,rf:muSR2,rf:muSR3,rf:muSR3-2,rf:muSR4}.

Recently, Katano {\it et al.}\cite{rf:Katano} and Lake {\it et al.}\cite{rf:Lake010} have found from the neutron scattering experiments in magnetic fields that the intensity of the incommensurate magnetic Bragg peak corresponding to the long-range static stripe order is enhanced around $p =$ 1/8 in La$_{2-x}$Sr$_x$CuO$_4$ by the application of magnetic fields parallel to the $c$-axis.
The enhancement of the magnetic Bragg peak intensity is marked in $x =$ 0.10\cite{rf:Lake010}, while it is observable but small in $x =$ 0.12\cite{rf:Katano}. 
A similar enhancement has also been found in the excess-oxygen-doped La$_2$CuO$_{4+\delta}$\cite{rf:Khaykovich}.
Such a field-induced magnetic order may be interpreted as being due to the pinning of the dynamical stripes by induced vortex cores in the $ab$-plane, though the pinning effect of vortex cores is not certain.
From the NMR measurements of the nearly optimally doped Tl$_2$Ba$_2$CuO$_{6+\delta}$, on the other hand, a magnetically ordered state has been detected only inside vortex cores, while the electronic state outside vortex cores remains superconducting\cite{rf:KumagaiNMR}.
This seems to be inconsistent with the above results of the neutron scattering experiments suggesting the development of the long-range magnetic order in magnetic fields.
As for the theoretical study on the field-induced magnetic order, before the report of these experimental results, the SO(5) theory had already predicted that the electronic state in vortex cores was an antiferromagnetically ordered one\cite{rf:Zhang,rf:Arovas}. 
According to the recent theoretical study by Demler {\it et al.}\cite{rf:Demler}, on the other hand, the results of the neutron scattering experiments in a magnetic field may be explained as being due to the approach of the superconducting phase to the coexisting phase of the superconducting state and the magnetically ordered state as a result of the increase of the spin fluctuations with low energy induced by the proximity effect. 
Accordingly, the origin of the field-induced magnetic order in the La-based high-$T_{\rm c}$ cuprates has not yet been settled.

The thermal conductivity measurement is a renewed technique for study of the spin and charge state in transition-metal oxides.
For example, it has been found that the thermal conductivity due to phonons is markedly enhanced in the spin-gap state of CuGeO$_3$\cite{rf:AndoCuGe} and SrCu$_2$(BO$_3$)$_2$\cite{rf:KudoSr,rf:Hof,rf:KudoSr2} and also in the charge-ordered state of Sr$_{1.5}$La$_{0.5}$MnO$_4$\cite{rf:Hess}.
The enhancement of the thermal conductivity due to phonons has also been reported in the static stripe-ordered state of La$_{1.28}$Nd$_{0.6}$Sr$_{0.12}$CuO$_4$\cite{rf:Baberski} and La$_{2-x}$Sr$_x$NiO$_4$\cite{rf:Hess}, while the suppression of the thermal conductivity due to phonons has been reported in the possible short-range (dynamical) stripe-ordered state of YBa$_2$Cu$_3$O$_{7-\delta}$ and HgBa$_2$Ca$_{n-1}$Cu$_n$O$_{2n+2+\delta}$ (n $=$ 1, 2, 3) around $p =$ 1/8\cite{rf:Cohn}.
In the antiferromagnetically ordered state of La$_2$CuO$_4$ \cite{rf:214-1,rf:214-2,rf:214-3} and YBa$_2$Cu$_3$O$_6$\cite{rf:123}, large contribution of magnons to the thermal conductivity in the $ab$-plane has been observed.

In this paper, in order to investigate the field-induced magnetically ordered state of La$_{2-x}$Sr$_x$CuO$_4$ around $p =$ 1/8, we have performed the thermal conductivity measurements in the $ab$-plane of La$_{2-x}$Sr$_x$CuO$_4$ ($x =$ 0.10, 0.115, 0.13) and 1 \% Zn-substituted La$_{2-x}$Sr$_x$Cu$_{0.99}$Zn$_{0.01}$O$_4$ ($x =$ 0.115) single crystals in magnetic fields parallel to the $c$-axis and also parallel to the $ab$-plane. 
This is because the suppression of the superconductivity occurs around $x =$ 0.115 in La$_{2-x}$Sr$_{x}$CuO$_4$ and because the suppression becomes most marked at $x =$ 0.115 through the partial substitution of Zn for Cu\cite{rf:Koike115}.

\section{experimental}
Single crystals were grown by the Traveling-Solvent Floating-Zone method in a similar way to that described in Ref. \cite{rf:Kawamata}.
Thermal conductivity measurements were carried out by a conventional steady-state method. 
One side of a rectangular single-crystal, whose typical dimensions were $3 \times 1 \times 1$ mm$^3$, was anchored on the copper heat sink with indium solder. 
A chip-resistance of 1 k$\Omega$ (Alpha Electronics Corp., MP1K000) was attached as a heater to the opposite side of the single crystal with GE7031 varnish. 
The temperature difference across the crystal (0.02--1.0 K) was measured with two Cernox thermometers (LakeShore Cryotronics, Inc., CX-1050-SD). 
Magnetic fields up to 14 T were applied parallel to the $c$-axis and also parallel to the $ab$-plane, using a superconducting magnet.

\section{Results and Discussion}
\begin{figure*}[hbt]
\begin{center}
\includegraphics[width=1\linewidth]{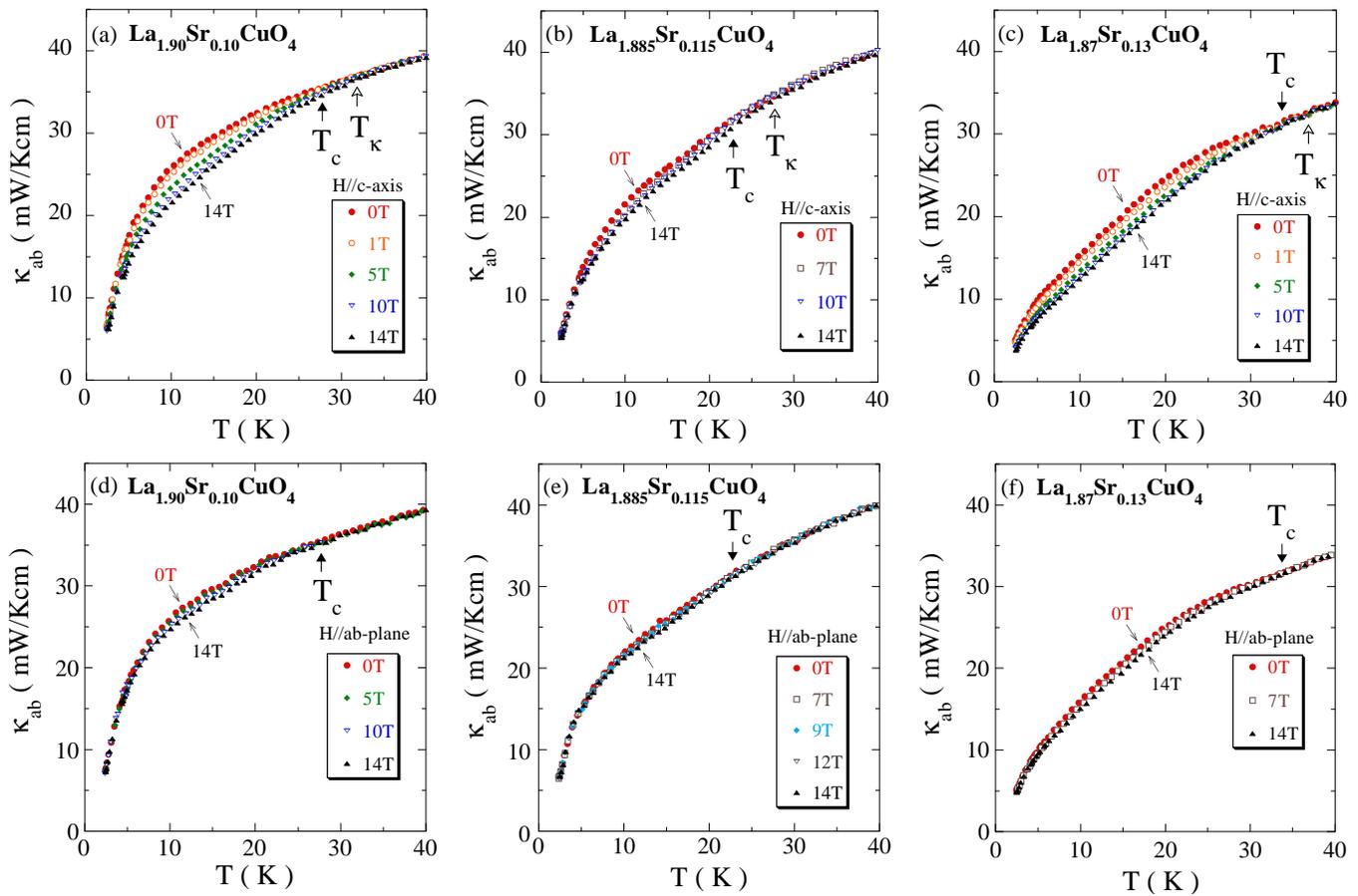}
\caption{(color online) Temperature dependence of the in-plane thermal conductivity, $\kappa_{\rm ab}$, of La$_{2-x}$Sr$_x$CuO$_4$ ($x =$ 0.10, 0.115, 0.13) in magnetic fields [(a)-(c)] parallel to the $c$-axis and [(d)-(f)] parallel to the $ab$-plane. 
Closed and open arrows denote the superconducting transition temperature $T_{\rm c}$ and the temperature $T_\kappa$ below which the thermal conductivity is suppressed by the application of magnetic fields, respectively.
}\label{fig:1}
\end{center}
\end{figure*}
Figures \ref{fig:1} (a)-(c) show the temperature dependence of the thermal conductivity in the $ab$-plane, $\kappa_{\rm ab}$, of La$_{2-x}$Sr$_x$CuO$_4$ ($x =$ 0.10, 0.115, 0.13) in magnetic fields parallel to the $c$-axis. 
As formerly reported by Nakamura {\it et al.}\cite{rf:214-1}, $\kappa_{\rm ab}$ in zero field decreases with decreasing temperature, and it exhibits a slight enhancement below the superconducting transition temperature $T_{\rm c}$, defined as the onset temperature of the Meissner diamagnetism. 
By the application of magnetic fields,  $\kappa_{\rm ab}$ is suppressed at low temperatures. 
Strictly looking, it appears that the temperature $T_\kappa$,  below which $\kappa_{\rm ab}$ is suppressed, is located above $T_{\rm c}$ and is almost independent of the magnitude of the magnetic field.
The suppression by the application of magnetic fields is marked in $x =$ 0.10 and 0.13, while it is small in $x =$ 0.115.
By the application of magnetic fields parallel to the $ab$-plane, on the other hand, the suppression is so small in all $x$ that $T_\kappa$ is hard to be determined, as shown in Figs. \ref{fig:1} (d)-(f). 
\begin{figure*}[htb]
\begin{center}
\includegraphics[width=1\linewidth]{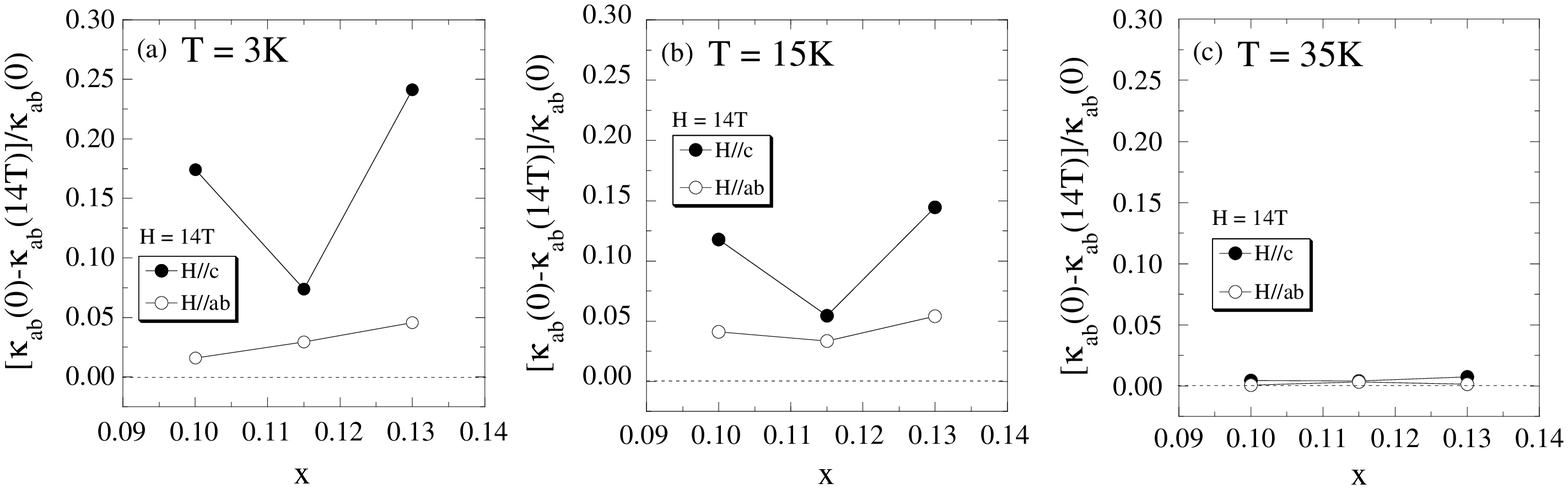}
\caption{Dependence on $x$ of the relative reduction of the in-plane thermal conductivity of La$_{2-x}$Sr$_x$CuO$_4$ ($x =$ 0.10, 0.115, 0.13), $[\kappa_{\rm ab}(0 {\rm T}) - \kappa_{\rm ab}(14 {\rm T})]/\kappa_{\rm ab}(0 {\rm T})$, in a magnetic field of 14 T parallel to the $c$-axis and also parallel to the $ab$-plane at (a) 3 K, (b) 15 K and (c) 35 K.}\label{fig:2}
\end{center}
\end{figure*}
Figures \ref{fig:2} (a)-(c) display the $x$ dependence of the relative reduction of $\kappa_{\rm ab}$, $[\kappa_{\rm ab}({\rm 0 T}) - \kappa_{\rm ab}({\rm 14 T})]/\kappa_{\rm ab}({\rm 0 T})$, in a magnetic field of 14 T parallel to the $c$-axis and also parallel to the $ab$-plane.
It is clearly seen that the relative reduction in a field parallel to the $c$-axis exhibits a dip at $x =$ 0.115 at low temperatures below $T_\kappa$, while that in the field parallel to the $ab$-plane is almost independent of $x$ in the whole temperature range. 
The former $x$ dependence with a dip-like shape is similar to the dip-like shape around $p =$ 1/8 in the $T_{\rm c}$ vs $x$ diagram, being associated with the 1/8 anomaly and the stripe order.

First, we try explaining the observed field-dependence of  $\kappa_{\rm ab}$ in terms of conventional mechanisms. 
It is well known that the thermal conductivity in the high-$T_{\rm c}$ cuprates exhibits an increase just below $T_{\rm c}$ with decreasing temperature, which is explained as being due to the increase of the thermal conductivity due to both phonons and electrons on account of the decrease of the phonon-electron scattering rate and the increase of the life time of quasiparticles, respectively\cite{rf:Hagen-ph,rf:Yu-ele}. 
By the application of magnetic fields, the thermal conductivity due to quasiparticles is expected to be enhanced in such d-wave superconductors as the high-$T_{\rm c}$ cuprates owing to the Volovik effect\cite{rf:Volovik}, namely, the increase of the density of states at the Fermi level, because the supercurrent surrounding vortices causes the Doppler shift of the energy of quasiparticles around the nodes of the superconducting gap\cite{rf:Sun1,rf:Izawa}. 
In detail, recent theories have suggested two kinds of pictures for the state of quasiparticles in magnetic fields. 
One is a group of discrete states induced by the Andreev reflection which are composed of localized states within vortex cores\cite{rf:Kopnin1,rf:Kopnin2} and quantized states expanding outside vortex cores\cite{rf:Anderson,rf:Gorkov,rf:Janko,rf:Kopnin3}. 
The other is an energy-band state due to the periodic vortex potential\cite{rf:Melnikov,rf:Franz}. 
In any case, however, these are contrary to the present experimental result. 
In order to understand the suppression of the thermal conductivity due to quasiparticles in magnetic fields, we should consider that the increase of quasiparticles leads to the increase of the electron-electron scattering rate and/or the increase of the electron-vortex scattering rate. 
The latter has been proposed to understand the field dependence of the in-plane thermal conductivity in Bi$_2$Sr$_2$CaCu$_2$O$_8$\cite{rf:Krishana,rf:Ando2212}. 
As for the thermal conductivity due to phonons, on the other hand, it is expected to be suppressed by the application of magnetic fields owing to the increase of the phonon-vortex scattering rate and/or the increase of the phonon-electron scattering rate, because the number of quasiparticles increases due to the Doppler shift. 
Hence, at a glance, the suppression of the thermal conductivity does not seem anomalous at all. 
However, these conventional mechanisms can not explain the fairly small field-dependence of $\kappa_{\rm ab}$ in $x =$ 0.115 compared with $\kappa_{\rm ab}$ in $x =$ 0.10 and 0.13.
Therefore, a new concept is necessary to explain the suppression of the thermal conductivity by the application of magnetic fields in La$_{2-x}$Sr$_x$CuO$_4$.

\begin{figure*}[htb]
\begin{center}
\includegraphics[width=1\linewidth]{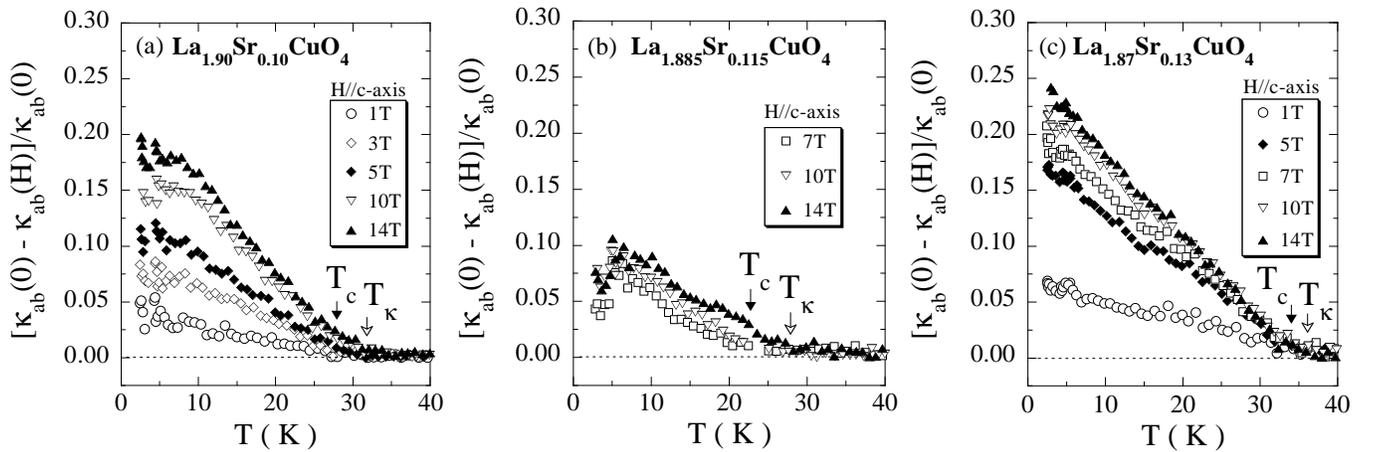}
\caption{Temperature dependence of the relative reduction of the in-plane thermal conductivity, $[\kappa_{\rm ab}(0 {\rm T}) - \kappa_{\rm ab}(H)]/\kappa_{\rm ab}(0 {\rm T})$, in several magnetic fields for La$_{2-x}$Sr$_x$CuO$_4$ with (a) $x =$ 0.10, (b) $x =$ 0.115 and (c) $x =$ 0.13. Closed and open arrows denote the superconducting transition temperature $T_{\rm c}$ and the temperature $T_\kappa$ below which the thermal conductivity is suppressed by the application of magnetic fields, respectively.}\label{fig:3}
\end{center}
\end{figure*}

Next, in order to explain the anomalous $x$ dependence of the suppression of the thermal conductivity in magnetic fields, we focus our mind on the development of the static stripe order in magnetic fields. 
In the following discussion, the field-induced magnetic order is regarded as the field-induced static stripe order. 
For comparison with the neutron scattering data, the temperature dependence of the relative reduction of the thermal conductivity, $[\kappa_{\rm ab}({\rm 0 T}) - \kappa_{\rm ab}(H)]/\kappa_{\rm ab}({\rm 0 T})$, in several magnetic fields $H$ parallel to the $c$-axis is plotted in Figs. \ref{fig:3}(a)-(c).
For $x =$ 0.10, it is found that the relative reduction rapidly increases at low temperatures below $T_\kappa$ with decreasing temperature and that $T_\kappa$ is almost independent of the magnitude of the magnetic field. 
The relative reduction is found to increase with increasing magnetic-field. 
These temperature and field dependences are quite similar to those of the intensity of the magnetic Bragg peak corresponding to the static stripe order in magnetic fields in $x =$ 0.10\cite{rf:Lake010}.
Moreover, $T_\kappa$ of $x =$ 0.10 coincides with the temperature $T_{\rm M}$ of $x =$ 0.10 below which the magnetic Bragg peak in magnetic fields develops\cite{rf:Lake010}. 
Therefore, it is naturally understood that the suppression of the thermal conductivity in magnetic fields is related to the development of the static stripe order.

As mentioned in Sec. I, there are some previous reports on the relation between the thermal conductivity due to phonons and the static stripe order. 
It has been reported that the thermal conductivity due to phonons is enhanced by the formation of the static stripe order for La$_{1.28}$Nd$_{0.6}$Sr$_{0.12}$CuO$_4$\cite{rf:Baberski} and La$_{2-x}$Sr$_x$NiO$_4$\cite{rf:Hess} and that, on the contrary, it is suppressed in the possible short-range (dynamical) stripe-ordered state of YBa$_2$Cu$_3$O$_{7-\delta}$ and HgBa$_2$Ca$_{n-1}$Cu$_n$O$_{2n+2+\delta}$ ($n =$ 1, 2, 3) around $p = 1/8$\cite{rf:Cohn}. 
The enhancement of the thermal conductivity due to phonons in La$_{1.28}$Nd$_{0.6}$Sr$_{0.12}$CuO$_4$ has been interpreted as being due to the disappearance of the scattering of phonons by the dynamical stripes which make the mean free path of phonons strongly limited\cite{rf:Baberski}. 
In this case, phonons are expected not to be scattered so strongly by the lattice distortion induced by the static stripe order in La$_{1.28}$Nd$_{0.6}$Sr$_{0.12}$CuO$_4$, because the correlation length of the static stripe order is more than $\sim 170 {\rm \AA}$ so that the lattice distortion is regarded as being rather periodic\cite{rf:Tranquada2}.
In the case of YBa$_2$Cu$_3$O$_{7-\delta}$, on the other hand, the suppression of the thermal conductivity due to phonons has been interpreted as being due to the possible formation of the short-range stripe order, namely, the dynamical stripe correlations, which was suggested from the inelastic neutron scattering experiment in YBa$_2$Cu$_3$O$_{6.6}$ by Mook {\it et al.}\cite{rf:Mook}. 
Accordingly, it is found to depend on the correlation length of the stripe order whether the thermal conductivity is enhanced or suppressed, but the present suppression of the thermal conductivity in magnetic fields in La$_{2-x}$Sr$_x$CuO$_4$ cannot be explained on these lines, because the correlation length of the static stripe order of $x =$ 0.10 in a magnetic field of 14.5 T\cite{rf:Lake010} is as long as that of La$_{1.48}$Nd$_{0.4}$Sr$_{0.12}$CuO$_4$ in zero field\cite{rf:Tranquada2}. 
According to the detailed investigation on the effect of the structural phase transition on the thermal conductivity for La$_{2-x-y}$Nd$_y$Sr$_x$CuO$_4$ by Sera {\it et al.}\cite{rf:Sera}, on the other hand, the enhancement of the thermal conductivity is explained as being due to the increase of the phonon velocity through the structural phase transition to the TLT structure. 
Very recently, moreover, it has been pointed out by Hess {\it et al.}\cite{rf:Hess3} that the enhancement in the TLT phase of La$_{2-x-y}$RE$_y$Sr$_x$CuO$_4$ (RE: rare earth element) is probably not due to the formation of the static stripe order but due to the suppression of the lattice instability. 
At present, therefore, it appears that the formation of the static stripe order does not have a large effect on the thermal conductivity due to phonons and it is hard to clearly explain the present suppression of the thermal conductivity in magnetic fields on the basis of the thermal conductivity due to phonons.

Alternatively, we try to understand the suppression of the thermal conductivity on the basis of the change of the thermal conductivity due to quasiparticles through the field-induced static stripe order. 
In the CuO$_2$ plane where the static stripe order is formed, charge carriers (holes) are confined in the one-dimensional path of the stripe\cite{rf:AndoR-T}, so that they cannot move so easily as in the carrier-homogeneous CuO$_2$ plane, leading to the decrease of the mobility of quasiparticles carrying heat. 
In fact, it has been pointed out from the electrical resistivity\cite{rf:Boe} and thermal conductivity\cite{rf:Sun1} measurements that quasiparticles tend to be localized by the application of magnetic fields in the underdoped region of La$_{2-x}$Sr$_x$CuO$_4$.
Moreover, when both the static stripe-ordered phase and the superconducting phase coexist and form domains, the mean free path of quasiparticles carrying heat is expected to be reduced by the domain wall. 
Accordingly, the suppression of the thermal conductivity in magnetic fields in La$_{2-x}$Sr$_x$CuO$_4$ is able to be ascribed to the decrease of the thermal conductivity due to quasiparticles through the development of the static stripe order, and its anomalous $x$ dependence is also explained as follows.
That is to say, the small field dependence in $x =$ 0.115 is attributed to the fact that the static stripe order is already developed in $x =$ 0.115 even in zero field\cite{rf:Suzuki,rf:Kimura2}, while the large field dependence in $x =$ 0.10 and 0.13 is able to be regarded as being due to the marked development of the static stripe order by the application of magnetic fields.
As shown in Figs. \ref{fig:3}(a)-(c), in fact, the field dependence of the relative reduction of the thermal conductivity is very small in $x =$ 0.115, compared with the large field-dependence in $x =$ 0.10 and 0.13.
Here, it is noted that the temperature dependence of the relative reduction of the thermal conductivity in $x =$ 0.115 and 0.13 roughly indicate the temperature dependence of the intensity of the magnetic Bragg peak corresponding to the static stripe order in magnetic fields, as in the case of $x =$ 0.10. 
Furthermore, it is worth while noting that Figs. \ref{fig:3}(a)-(c) perhaps represent the development of the charge stripe order in $x =$ 0.10, 0.115 and 0.13.
By the way, this model gives a possible answer for the well-known question why the enhancement of the thermal conductivity at low temperatures below $T_{\rm c}$ is relatively small in La$_{\rm 2-x}$Sr$_x$CuO$_4$\cite{rf:214-1} compared with the other high-$T_{\rm c}$ cuprates\cite{rf:Yu-ele,rf:Hagen-ph}. 
In the underdoped region, this is because at least a small amount of the static stripe order exists even in zero field to strongly suppress the thermal conductivity, though the magnetic Bragg peak detected in the neutron scattering experiments is very weak\cite{rf:Suzuki,rf:Kimura2,rf:Matsushita}.

\begin{figure}[tb]
\begin{center}
\includegraphics[width=0.7\linewidth]{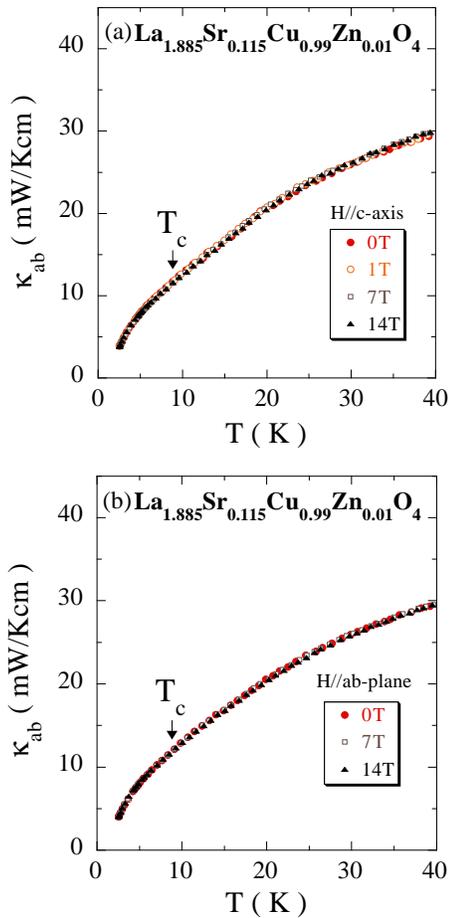}
\caption{(color online) Temperature dependence of the in-plane thermal conductivity, $\kappa_{\rm ab}$, of the 1 \% Zn-substituted La$_{2-x}$Sr$_x$Cu$_{0.99}$Zn$_{0.01}$O$_4$ ($x =$ 0.115) in magnetic fields (a) parallel to the $c$-axis and (b) parallel to the $ab$-plane. Arrows denote the superconducting temperature $T_{\rm c}$.}\label{fig:4}
\end{center}
\end{figure}

The above model is confirmed by the following experimental result for the 1 \% Zn-substituted La$_{2-x}$Sr$_x$Cu$_{0.99}$Zn$_{0.01}$O$_4$ with $x =$ 0.115, as shown in Fig. \ref{fig:4}. 
It is found that the thermal conductivity exhibits no field-dependence in magnetic fields parallel to the $c$-axis and also parallel to the $ab$-plane. 
According to the above model, no field-dependence is reasonably understood, because the static stripe order is fully developed even in zero field on account of the  strong pinning effect of Zn\cite{rf:Koike0,rf:Koike,rf:Adachi,rf:muSR0,rf:muSR1,rf:muSR2,rf:muSR3,rf:muSR3-2,rf:muSR4} so that further development of the static stripe order by the application of magnetic fields is not expected.

Now, there remains a question why $T_\kappa$ is almost independent of the magnitude of the magnetic field, though this question has already been pointed out from the neutron scattering experiment\cite{rf:Lake010}. 
Before answering the question, first we suppose the stripe pinning model by vortex cores. 
Taking into account the result that the dynamical stripes of spins and holes are pinned in such an inhomogeneous electronic background as the partially Zn-substituted CuO$_2$ plane\cite{rf:Koike0,rf:Koike,rf:Adachi,rf:muSR0,rf:muSR1,rf:muSR2,rf:muSR3,rf:muSR3-2,rf:muSR4,rf:Smith,rf:Tohyama}, vortex cores induced in the CuO$_2$ plane are also expected to operate to pin the dynamical stripes, as mentioned in Sec. I. 
Surely, the existence of impurities such as Zn atoms or vortex cores induces the energy loss in the dynamical stripes, but the most important point is that the energy loss depends on whether an impurity such as Zn or a vortex core is located at a charge stripe or a spin stripe. 
The stripe order should be located in a way which make the energy loss the smallest, so that the dynamical stripes are pinned, resulting in the development of the static stripe order. 
The stripe pinning model by vortex cores is supported by the present experimental result that the suppression of the thermal conductivity is not observed by the application of magnetic fields parallel to the $ab$-plane. 
This is because, in magnetic fields parallel to the $ab$-plane, vortex cores penetrate the so-called blocking-layer preferably, so that no vortex core appears in the CuO$_2$ plane, leading to neither pinning of the dynamical stripes nor development of the static stripe order. 
Recurring to the question, the distance between vortex cores is $\sim$ 130 ${\rm \AA}$ in a magnetic field of 14 T. 
Therefore, the pinning effect of vortex cores seems to be so local that $T_\kappa$ does not depend on the number of vortex cores at least in magnetic fields up to 14 T. 
Meanwhile, it is found that $T_\kappa$ is located above $T_{\rm c}$, but this is not inconsistent with the above discussion, because the superconducting fluctuation exists above $T_{\rm c}$.
In fact, recent reports on the Nernst effect have suggested that the vortex state survives even above $T_{\rm c}$\cite{rf:Xu,rf:Ong} . 
Thus, the field independence of $T_\kappa$ is explained in terms of the stripe pinning model by vortex cores\cite{rf:footnote}.

\begin{figure}[htb]
\begin{center}
\includegraphics[width=0.8\linewidth]{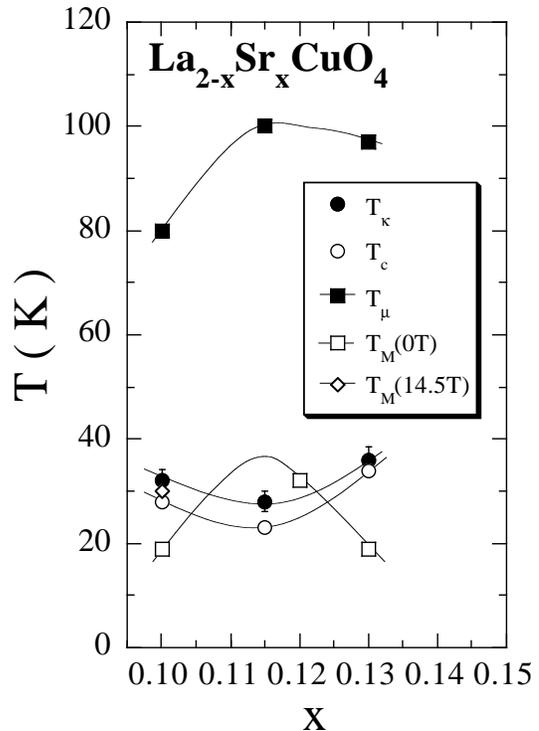}
\caption{Variations with $x$ of the temperature $T_\kappa$ below which the in-plane thermal conductivity is suppressed by the application of magnetic fields parallel to the $c$-axis, the superconducting transition temperature $T_{\rm c}$, the temperature $T_\mu$ below which the development of the dynamical spin correlation was detected in zero field from the $\mu$SR measurements\cite{rf:highTG0,rf:highTG} and the temperatures $T_{\rm M}$(0)\cite{rf:Suzuki,rf:Kimura2,rf:Matsushita} and $T_{\rm M}$(14.5 T)\cite{rf:Lake010} below which the magnetic Bragg peak corresponding to the static stripe order develops in the neutron scattering measurements in zero field and in a magnetic field of 14.5 T parallel to the $c$-axis, respectively. Solid lines are guides for eyes.
}\label{fig:5}
\end{center}
\end{figure}
Finally, in order to examine the stripe pinning model by vortex cores, we compare the present results with the neutron scattering results\cite{rf:Lake010,rf:Suzuki,rf:Kimura2,rf:Matsushita} and  also the recent $\mu$SR results revealing a peculiar $x$ dependence of the temperature $T_\mu$ below which the development of the dynamical spin correlations was detected in zero field from the $\mu$SR measurements\cite{rf:highTG0,rf:highTG}. 
Figure \ref{fig:5} shows $x$ dependences of $T_\kappa$, $T_{\rm c}$, $T_\mu$, $T_{\rm M}$(0) in zero field and $T_{\rm M}$(14.5 T) in a magnetic field of 14.5 T parallel to the $c$-axis.
It is found that both $T_{\rm M}$(0) and $T_\mu$ exhibit the maximum at $x \sim$ 0.115, while $T_\kappa$ exhibits the minimum at $x =$ 0.115. 
That is, the $x$ dependences of $T_{\rm M}$(0) and $T_\mu$ are contrary to that of $T_\kappa$, so that $T_\kappa$ is directly correlated with $T_{\rm c}$ rather than $T_{\rm M}$(0) and $T_\mu$.
This suggests that $T_\kappa$ is regarded as the temperature below which vortex cores are formed in the superconducting fluctuation region in the CuO$_2$ plane so as to pin the dynamical stripes and develop the static stripe order.
In fact, the observed field dependence of the thermal conductivity is well explained as follows. 
For $x =$ 0.115, $T_\kappa$ is lower than $T_{\rm M}$(0), so that further development of the static stripe order is not large at low temperatures below $T_\kappa$, leading to the fairly small field-dependence of the thermal conductivity. 
For $x =$ 0.10 and 0.13, on the other hand, $T_\kappa$ is much higher than $T_{\rm M}$(0), so that the development of the static stripe order is marked at low temperatures below $T_\kappa$, leading to the large field-dependence of the thermal conductivity. 
The result that $T_\kappa$ is much higher than $T_{\rm M}$(0) for $x =$ 0.10 and 0.13 is analogous to the result obtained from the $\mu$SR measurements that the long-range magnetic ordering temperature $T_{\rm N}$ is much higher in the lightly Zn-substituted La$_{2-x}$Sr$_x$Cu$_{1-y}$Zn$_y$O$_4$ than in the Zn-free La$_{2-x}$Sr$_x$CuO$_4$ for $x =$ 0.10 and 0.13\cite{rf:muSR1,rf:muSR2,rf:muSR3,rf:muSR3-2,rf:muSR4}, suggesting that vortex cores operate to pin the dynamical stripes as Zn atoms.
This is consistent with the fact that $T_{\rm M}$(14.5 T) deviates from $T_{\rm M}$(0) and roughly coincides with $T_\kappa$.
Accordingly, the present results strongly suggest that the origin of the field-induced magnetic order is the pinning of the dynamical stripes by vortex cores.

\section{Conclusions}
 We have measured the thermal conductivity in the $ab$-plane of La$_{2-x}$Sr$_x$CuO$_4$ ($x =$ 0.10, 0.115, 0.13) in magnetic fields up to 14 T parallel to the $c$-axis and also parallel to the $ab$-plane.
By the application of magnetic fields parallel to the $c$-axis, the thermal conductivity has been found to be suppressed at low temperatures below the temperature $T_\kappa$ which is located above $T_{\rm c}$ and almost independent of the magnitude of the magnetic field. 
The suppression is marked in $x =$ 0.10 and 0.13, while it is small in $x =$ 0.115. 
Furthermore, no suppression is observed in the 1 \% Zn-substituted La$_{2-x}$Sr$_x$Cu$_{0.99}$Zn$_{0.01}$O$_4$ with $x =$ 0.115. 
Taking into account the experimental results that the temperature dependence of the relative reduction of the thermal conductivity in $x =$ 0.10 is quite similar to the temperature dependence of the intensity of the magnetic Bragg peak corresponding to the static stripe order in magnetic fields and that the Zn substitution tends to stabilize the static order, it is concluded that the suppression of the thermal conductivity in magnetic fields is attributed to the development of the static stripe order. 
The temperature dependence of the relative reduction of the thermal conductivity in $x =$ 0.115 and 0.13 may indicate the temperature dependence of the intensity of the magnetic Bragg peak corresponding to the static stripe order in magnetic fields, as in the case of $x =$ 0.10. 
Moreover, it has been found that $T_\kappa$ is directly correlated with $T_{\rm c}$ rather than $T_{\rm M}$(0) and $T_\mu$, so that $T_\kappa$ is regarded as the temperature below which vortex cores pin the dynamical stripes of spins and holes.
The present results suggest that the field-induced magnetic order in La$_{2-x}$Sr$_x$CuO$_4$ originates from the pinning of the dynamical stripes of spins and holes by vortex cores.

\acknowledgements
We are grateful to Professor T. Tohyama and Doctor M. Fujita for the useful discussion.
The thermal conductivity measurements were performed at the High Field Laboratory for Superconducting Materials, Institute for Materials Research, Tohoku University.
This work was supported by a Grant-in Aid for Scientific Research from the Ministry of Education, Culture, Sports, Science and Technology, Japan.
One of the authors (K. K.) was supported by the Japan Society for the Promotion of Science.


\begin{references}
\bibitem{rf:Tranquada} J. M. Tranquada, B. J. Sternlieb, J. D. Axe, Y. Nakamura, and S. Uchida, Nature (London) {\bf 375}, 561 (1995). 
\bibitem{rf:Cheong} S. -W. Cheong, G. Aeppli, T. E. Mason, H. Mook, S. M. Hayden, P. C. Canfield, Z. Fisk, K. N. Clausen, and J. L. Martinez, Phys. Rev. Lett. {\bf 67}, 1791 (1991).
\bibitem{rf:Mason} T. E. Mason, G. Aeppli, and H. A. Mook, Phys. Rev. Lett. {\bf 68}, 1414 (1992). 
\bibitem{rf:Thurston} T. R. Thurston, P. M. Gehring, G. Shirane, R. J. Birgeneau, M. A. Kastner, Y. Endoh, M. Matsuda, K. Yamada, H. Kojima, and I. Tanaka, Phys. Rev. B {\bf 46}, 9128 (1992). 
\bibitem{rf:Yamada} K. Yamada, C. H. Lee, K. Kurahashi, J. Wada, S. Wakimoto, S. Ueki, H. Kimura, Y. Endoh, S. Hosoya, G. Shirane, R. J. Birgeneau, M. Greven, M. A. Kastner, and Y. J. Kim, Phys. Rev. B {\bf 57}, 6165 (1998).
\bibitem{rf:Kumagai} K. Kumagai, I. Watanabe, K. Kawano, H. Matoba, K. Nishiyama, K. Nagamine, N. Wada, M. Okaji, and K. Nara, Physica C {\bf 185-189}, 913 (1991). 
\bibitem{rf:Luke} G. M. Luke, L. P. Le, B. J. Sternlieb, W. D. Wu, Y. J. Uemura, J. H. Brewer, T. M. Riseman, S. Ishibashi, and S. Uchida, Physica C {\bf 185-189}, 1175 (1991). 
\bibitem{rf:Watanabe} I. Watanabe, K. Kawano, K. Kumagai, K. Nishiyama, and K. Nagamine, J. Phys. Soc. Jpn. {\bf 61}, 3058 (1992). 
\bibitem{rf:Watanabe2} I. Watanabe, K. Nishiyama, K. Nagamine, K. Kawano, and K. Kumagai, Hyperfine Interact. {\bf 86}, 603 (1994). 
\bibitem{rf:Koike0} Y. Koike, S. Takeuchi, H. Sato, Y. Hama, M. Kato, Y. Ono, and S. Katano, J. Low Temp. Phys. {\bf 105}, 317 (1996).
\bibitem{rf:Koike} Y. Koike, S. Takeuchi, Y. Hama, H. Sato, T. Adachi, and M. Kato, Physica C {\bf 282-287}, 1233 (1997). 
\bibitem{rf:Adachi} T. Adachi, T. Noji, H. Sato, Y. Koike, T. Nishizaki, and N. Kobayashi, J. Low Temp. Phys. {\bf 117}, 1151 (1999). 
\bibitem{rf:muSR0} I. Watanabe, and K. Nagamine, Physica B {\bf 259-261}, 544 (1999).
\bibitem{rf:muSR1} I. Watanabe, T. Adachi, K. Takahashi, S. Yairi, Y. Koike, and K. Nagamine, J. Phys. Chem. Solids {\bf 63}, 1093 (2002).
\bibitem{rf:muSR2} I. Watanabe, T. Adachi, K. Takahashi, S. Yairi, Y. Koike, and K. Nagamine, Phys. Rev. B {\bf 65}, 180516(R) (2002).
\bibitem{rf:muSR3} I. Watanabe, T. Adachi, S. Yairi, Y. Koike, and K. Nagamine, Physica B {\bf 326}, 305 (2003).
\bibitem{rf:muSR3-2} T. Adachi, I. Watanabe, S. Yairi, K. Takahashi, Y. Koike, and K. Nagamine, J. Low Temp. Phys. {\bf 131}, 843 (2003).
\bibitem{rf:muSR4} T. Adachi, S. Yairi, K. Takahashi, Y. Koike, I. Watanabe, and K. Nagamine, cond-mat/0306233. 
\bibitem{rf:Katano} S. Katano, M. Sato, K. Yamada, T. Suzuki, and T. Fukase, Phys. Rev. B {\bf 62}, R14677 (2000).
\bibitem{rf:Lake010} B. Lake, H. M. Ronnow, N. B. Christensen, G. Aeppli, K. Lefmann, D. F. McMorrow, P. Vorderwisch, P. Smeibidl, N. Mangkorntong, T. Sasagawa, M. Nohara, H. Takagi, and T. E. Mason, Nature (London) {\bf 415}, 299 (2002).
\bibitem{rf:Khaykovich} B. Khaykovich, Y. S. Lee, R. Erwin, S. -H. Lee, S. Wakimoto, K. J. Thomas, M. A. Kastner, and R. J. Birgeneau, Phys. Rev. B {\bf 66}, 014528 (2002).
\bibitem{rf:KumagaiNMR} K. Kakuyanagi, K. Kumagai, Y. Matsuda, and M. Hasegawa, Phys. Rev. Lett. {\bf 90}, 197003 (2003).
\bibitem{rf:Zhang} S. C. Zhang, Science {\bf 275}, 1089 (1997). 
\bibitem{rf:Arovas} D. P. Arovas, A. J. Berlinsky, C. Kallin, and S. -C. Zhang, Phys. Rev. Lett. {\bf 79}, 2871 (1997).
\bibitem{rf:Demler} E. Demler, S. Sachdev, and Y. Zhang, Phys. Rev. Lett. {\bf 87}, 067202 (2001). 
\bibitem{rf:AndoCuGe} Y. Ando, J. Takeya, D. L. Sisson, S. G. Doettinger, I. Tanaka, R. S. Feigelson, and A. Kapitulnik, Phys. Rev. B {\bf 58}, R2913 (1998). 
\bibitem{rf:KudoSr} K. Kudo, T. Noji, Y. Koike, T. Nishizaki, and N. Kobayashi, J. Phys. Soc. Jpn. {\bf 70}, 1448 (2001).
\bibitem{rf:Hof} M. Hofmann, T. Lorenz, G. S. Uhrig, H. Kierspel, O. Zabara, A. Freimuth, H. Kageyama, and Y. Ueda, Phys. Rev. Lett. {\bf 87}, 047202 (2001).
\bibitem{rf:KudoSr2} K. Kudo, T. Noji, Y. Koike, T. Nishizaki, and N. Kobayashi, Physica B {\bf 329-333}, 910 (2003). 
\bibitem{rf:Hess} C. Hess, B. B\"{u}chner, M. H\"{u}cker, R. Gross, and S. -W. Cheong, Phys. Rev. B {\bf 59}, R10397 (1999).
\bibitem{rf:Baberski} O. Baberski, A. Lang, O. Maldonado, M. H\"{u}cker, B. B\"{u}chner, and A. Freimuth, Europhys. Lett. {\bf 44}, 335 (1998).
\bibitem{rf:Cohn} J. L. Cohn, C. P. Popoviciu, Q. M. Lin, and C. W. Chu, Phys. Rev. B {\bf 59}, 3823 (1999).
\bibitem{rf:214-1} Y. Nakamura, S. Uchida, T. Kimura, M. Motohira, K. Kishio, K. Kitazawa, T. Arima, and Y. Tokura, Physica C {\bf 185-189}, 1409 (1991).
\bibitem{rf:214-2} X. F. Sun, J. Takeya, S. Komiya, and Y. Ando, Phys. Rev. B {\bf 67}, 104503 (2003). 
\bibitem{rf:214-3} C. Hess, B. B\"{u}chner, U. Ammerahl, L. Colonescu, F. Heidrich-Meisner, W. Brenig, and A. Revcolevschi, Phys. Rev. Lett. {\bf 90}, 197002 (2003).
\bibitem{rf:123} K. Takenaka, Y. Fukuzumi, K. Mizuhashi, S. Uchida, H. Asaoka, and H. Takei, Phys. Rev. B {\bf 56}, 5654 (1997). 
\bibitem{rf:Koike115} Y. Koike, A. Kobayashi, T. Kawaguchi, M. Kato, T. Noji, Y. Ono, T. Hikita and Y. Saito, Solid State Commun. {\bf 82}, 889 (1992). 
\bibitem{rf:Kawamata} T. Kawamata, T. Adachi, T. Noji, and Y. Koike, Phys. Rev. B {\bf 62}, R11981 (2000). 
\bibitem{rf:Yu-ele} R. C. Yu, M. B. Salamon, J. P. Lu, and W. C. Lee, Phys. Rev. Lett. {\bf 69}, 1431 (1992).
\bibitem{rf:Hagen-ph} S. J. Hagen, Z. Z. Wang, and N. P. Ong, Phys. Rev. B {\bf 40}, 9389 (1989).
\bibitem{rf:Volovik} G. E. Volovik, JETP Lett. {\bf 58}, 469 (1993). 
\bibitem{rf:Sun1} X. F. Sun, S. Komiya, J. Takeya, and Y. Ando, Phys. Rev. Lett. {\bf 90}, 117004 (2003).
\bibitem{rf:Izawa} K. Izawa, K. Kamata, Y. Nakajima, Y. Matsuda, T. Watanabe, M. Nohara, H. Takagi, P. Thalmeier, and K. Maki, Phys. Rev. Lett. {\bf 89}, 137006 (2002).
\bibitem{rf:Kopnin1} N. B. Kopnin and G. E. Volovik, Phys. Rev. Lett. 79, 1377 (1997). 
\bibitem{rf:Kopnin2} N. B. Kopnin, Phys. Rev. B {\bf 57}, 11775 (1998).
\bibitem{rf:Anderson} P. W. Anderson, cond-mat/9812063.
\bibitem{rf:Gorkov} L. P. Gor'kov and J. R. Schrieffer, Phys. Rev. Lett. {\bf 80}, 3360 (1998).
\bibitem{rf:Janko} B. Jank\'{o}, Phys. Rev. Lett. {\bf 82}, 4703 (1999). 
\bibitem{rf:Kopnin3} N. B. Kopnin and V. M. Vinokur, Phys. Rev. B {\bf 62}, 9770 (2000). 
\bibitem{rf:Melnikov} A. S. Mel'nikov, J. Phys.: Condens. Matter {\bf 11}, 4219 (1999). 
\bibitem{rf:Franz} M. Franz and Z. Te\v{s}anovi\'{c}, Phys. Rev. Lett. {\bf 84}, 554 (2000). 
\bibitem{rf:Krishana} K. Krishana, N. P. Ong, Q. Li, G. D. Gu, N. Koshizuka, Science {\bf 277}, 83 (1997). 
\bibitem{rf:Ando2212} Y. Ando, J. Takeya, Y. Abe, K. Nakamura and A. Kapitulnik, Phys. Rev. B {\bf 62}, 626 (2000). 
\bibitem{rf:Tranquada2} J. M. Tranquada, J. D. Axe, N. Ichikawa, Y. Nakamura, S. Uchida, and B. Nachumi, Phys. Rev. B {\bf 54}, 7489 (1996). 
\bibitem{rf:Mook} H. A. Mook, P. Dai, S. M. Hayden, G. Aeppli, T. G. Perring, and F. Do\u{g}an, Nature (London) {\bf 395}, 580 (1998).
\bibitem{rf:Sera} M. Sera, M. Maki, M. Hiroi, and N. Kobayashi, J. Phys. Soc. Jpn {\bf 66}, 765 (1997).
\bibitem{rf:Hess3} C. Hess, B. B\"{u}chner, U. Ammerahl, and A. Revcolevschi, cond-mat/0305321. 
\bibitem{rf:AndoR-T} Y. Ando, K. Segawa, S. Komiya, and A. N. Lavrov, Phys. Rev. Lett. {\bf 88}, 137005 (2002).
\bibitem{rf:Boe} G. S. Boebinger, Y. Ando, A. Passner, T. Kimura, M. Okuya, J. Shimoyama, K. Kishio, K. Tamasaku, N. Ichikawa, and S. Uchida, Phys. Rev. Lett. {\bf 77}, 5417 (1996).
\bibitem{rf:Suzuki} T. Suzuki, T. Goto, K. Chiba, T. Shinoda, T. Fukase, H. Kimura, K. Yamada, M. Ohashi, and Y. Yamaguchi, Phys. Rev. B {\bf 57}, 3229 (1998).
\bibitem{rf:Kimura2} H. Kimura, K. Hirota, H. Matsushita, K. Yamada, Y. Endoh, S. H. Lee, C. F. Majkrzak, R. Erwin, G. Shirane, M. Greven, Y. S. Lee, M. A. Kastner, and R. J. Birgeneau, Phys. Rev. B {\bf 59}, 6517 (1999).
\bibitem{rf:Matsushita} H. Matsushita, H. Kimura, M. Fujita, K. Yamada, K. Hirota, and Y. Endoh, J. Phys. Chem. Solids {\bf 60}, 1071 (1999).
\bibitem{rf:Smith} C. M. Smith, A. H. Castro Neto, and A. V. Balatsky, Phys. Rev. Lett. {\bf 87}, 177010 (2001).
\bibitem{rf:Tohyama} T. Tohyama, M. Takahashi, and S. Maekawa, Physica C {\bf 357-360}, 93 (2001).
\bibitem{rf:Xu} Z.A. Xu, N.P. Ong, Y. Wang, T. Kakeshita, and S. Uchida, Nature (London) {\bf 406}, 486 (2000).
\bibitem{rf:Ong} Y. Wang, Z. A. Xu, T. Kakeshita, S. Uchida, S. Ono, Y. Ando, and N. P. Ong, Phys. Rev. B {\bf 64}, 224519 (2001). 
\bibitem{rf:footnote} 
Here, one may wonder why $T_\kappa$ is independent of the magnitude of magnetic field if it is due to the onset of the superconducting fluctuations. 
However, taking into account that contour lines of the Nernst signal in the $H$ vs $T$ diagram rise very steeply around $T_{\rm c}$\cite{rf:Ong2}, the suppression of the onset temperature of the superconducting fluctuations by the application of  magnetic fields up to 15 T is considered to be too small to be detected within the accuracy of the present measurements. The field-independent $T_{\rm M}$ in the neutron scattering experiments is probably due to same reason.
\bibitem{rf:Ong2} Y. Wang, N. P. Ong, Z. A. Xu, T. Kakeshita, S. Uchida, D. A. Bonn, R. Liang, and W. N. Hardy, Phys. Rev. Lett. {\bf 88}, 257003 (2002).
\bibitem{rf:highTG0} I. Watanabe, T. Adachi, S. Yairi, H. Mikuni, Y. Koike, and K. Nagamine, J. Low Temp. Phys. {\bf 131}, 331 (2003).
\bibitem{rf:highTG} I. Watanabe, T. Adachi, S. Yairi, Y. Koike, and K. Nagamine, J. Magn. Magn. Mater (in press).
\end{references}


\end{document}